# Simulation studies of CZT(S,Se) single and tandem junction solar cells towards possibilities for higher efficiencies up to 22%

Goutam Kumar Gupta[1] and Ambesh Dixit[1]

[1]*Department of Physics and Center for Solar Energy, Indian institute of technology Jodhpur, Rajasthan, 342037, India*

*Corresponding author email:* *ambesh@iitj.ac.in*

## Abstract

We simulated photovoltaic characteristics of single heterojunction solar cell with $Cu_2ZnSnS_4$ and $Cu_2ZnSnSe_4$ absorber layer numerically using one dimensional solar cell capacitance simulator (SCAPS-1D). n-CdS/ZnO double buffer layer is used for hetrostructure interfaces with the absorber layer. The cell performance is investigated against variation of different material layer properties such as thickness, carrier concentration, defect density etc. The performance is optimized first for the single junction solar cell with Mo as back contact material with work function 5 eV. A double junction CZTS/CZTSe tandem cell structure is realized keeping the same material properties as is used in the single CZTS and CZTSe solar cell simulation and considering the flat band condition at the interface. Tandem cell performance is determined after matching the current condition for top and bottom sub-cells. The CZTS/CZTSe short circuit current density is ~ 20.98 mA/cm$^2$ for current matched *211.33 nm* thick CZTS top cell in conjunction with *2000 nm* bottom cell. The maximum efficiency obtained under the flat band condition at the contact is ~21.7% with open circuit voltage ~1.324 V.

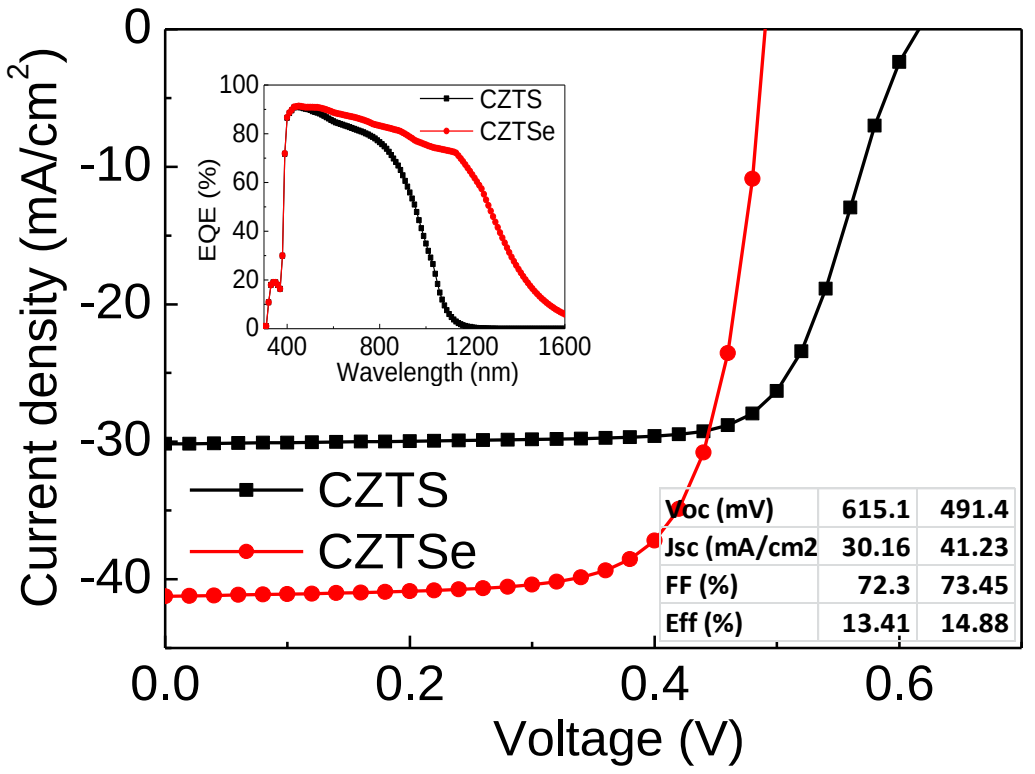

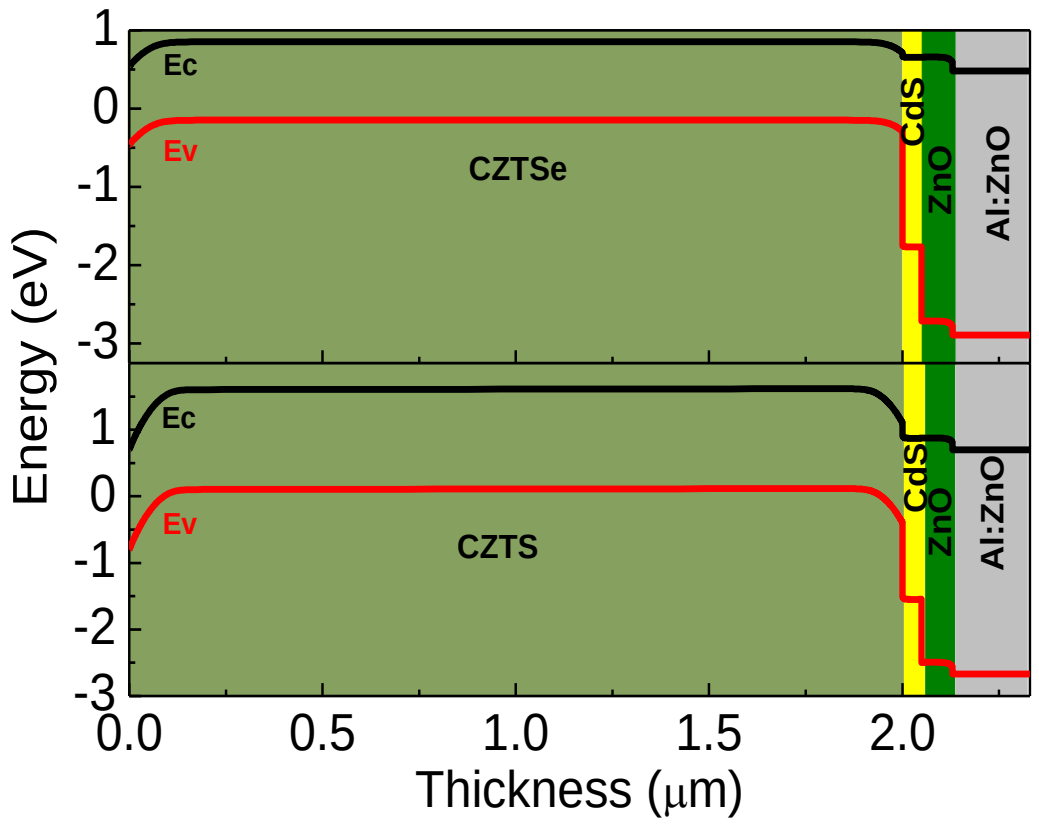

## Introduction:

Kesterite compound semiconductors $Cu_2ZnSn(S_xSe_{1-x})_4$ (CZTSSe) received considerable attention in the last decade due to its potential as a future absorber material for high efficiency thin film solar cell technology. The properties such as environmentally benign and earth abundant constituent elements, optimum direct bandgap (1-1.5 eV) and high absorption coefficient make CZTSSe a material of choice as an alternative to the existing solar photovoltaic thin film technologies, which are expected to scale upto several GW/year [1]. Considerable developments are reported towards improvement in the photovoltaic performance of CZTSSe kesterite based solar cells with the maximum efficiency ~ 9.2% for pure sulfide (CZTS) [2] and 11.6% for pure selenide (CZTSe) [3] based kesterite absorbers. The best kesterite cell efficiency reported till date is ~ 12.7% for sulfoselenide (CZTSSe) based absorber with double CdS/$In_2S_3$ as n type buffer layer and 12.6 % with single CdS buffer layer [4][5]. However, in spite of all these improvements, the maximum efficiency achieved for CZTSSe kesterite based absorbers is far below chalcopyrite $CuInGaSe_2$ (CIGS)[6] and CdTe based solar cells[7].

The similar crystal structure and material properties of kesterite provide the possibility of improvement in photovoltaic conversion efficiency. Al:ZnO/i-ZnO/CdS/CZTS,Se/Mo/SLG kesterite solar cell device structure is adopted from its chalcopyrite (CIGS) counterpart owing to its similar crystal structure and electronic properties[8]. The high short circuit current density with low open circuit voltage is always a trade-off with low energy band gap photovoltaic materials in a single junction configuration. Thus, it becomes essential to utilize the improved device structures by implementing the tandem structures (CZTS,Se) kesterite absorber to achieve the enhanced photovoltaic response for such material (CZTS,Se) material systems.

The tandem or multi-junction solar cell structures utilizes stacking of different band gap p-n junction solar cells in a specific configuration where the top cell should absorb the high energy portion of the solar spectrum, equivalent to absorber band gap, and the lower cells should absorb the part of the remaining spectrum, equivalent to the respective absorber band gaps. This tandem structure concept is well established with thin film solar cells, based on group III-V nitride based compound semiconductors and are used for space applications[9].

Recently, CIGS based stacked/tandem solar cell devices are explored, where $CuGaSe_2$ (CGS) absorber based solar cell is used as the top cell (Eg ~ 1.7 eV) and CIGS absorber cell has been used as the bottom cell (Eg ~1.14eV). Such CIGS tandem solar cell devices showed improved efficiency up to 7.2% with short circuit current 10.6 mA/cm$^2$ and an open-circuit voltage of 1.18V [10]. A similar silver and indium modified tandem solar cell structures e.g. $Ag(In_{0.2}Ga_{0.8})Se_2$ (AIGS) as top cell (Eg ~ 1.7 eV) and CIGS based solar cell as the bottom cell has also been reported with maximum efficiency, Jsc, and Voc values ~ 8%, 9.1mA/cm$^2$ and 1.3V, respectively [11].

In this work, we have considered tandem structure using CZTS/Se based absorber materials. The one dimensional electrical solar cell simulation program SCAPS-1D is used for simulating the device performances [15]. The desired material parameters, required for the present simulation studies are either borrowed from literature, cited wherever used or assumed with rationale for better understanding of the device performance under realistic situations. Initially, a single junction solar cell is optimized for both CZTS and CZTSe absorber layers independently and the effect of absorber layer thickness, carrier concentration and density of various defects are investigated. The studies are extended further for the tandem solar cell structures for realizing the enhanced photo conversion efficiencies. The proposed tandem solar cell structures based on CZTS/Se absorber material have shown photo conversion efficiency upto ~22%.

**Materials' properties and simulation approach:**

The considered device structure, shown in Fig 1 (a), includes CZTS and CZTSe as the p- type absorber layer, CdS as n-type wide band gap buffer layer, ZnO as a window layer, also acting as a passivation layer and Al doped ZnO (Al:ZnO) as a transparent conducting oxide (TCO) layer. The material properties are summarized in Table 1, used for simulating the photovoltaic response for the proposed single and tandem device structures. The optical absorption spectra for the CZTS and CZTSe are taken from literature[12] and for CdS and ZnO the absorption spectrum file as provided in SCAPS are used and are shown in Fig 1 (b). These absorption data for different layers are used in the present work to mimic the practical absorbing properties.

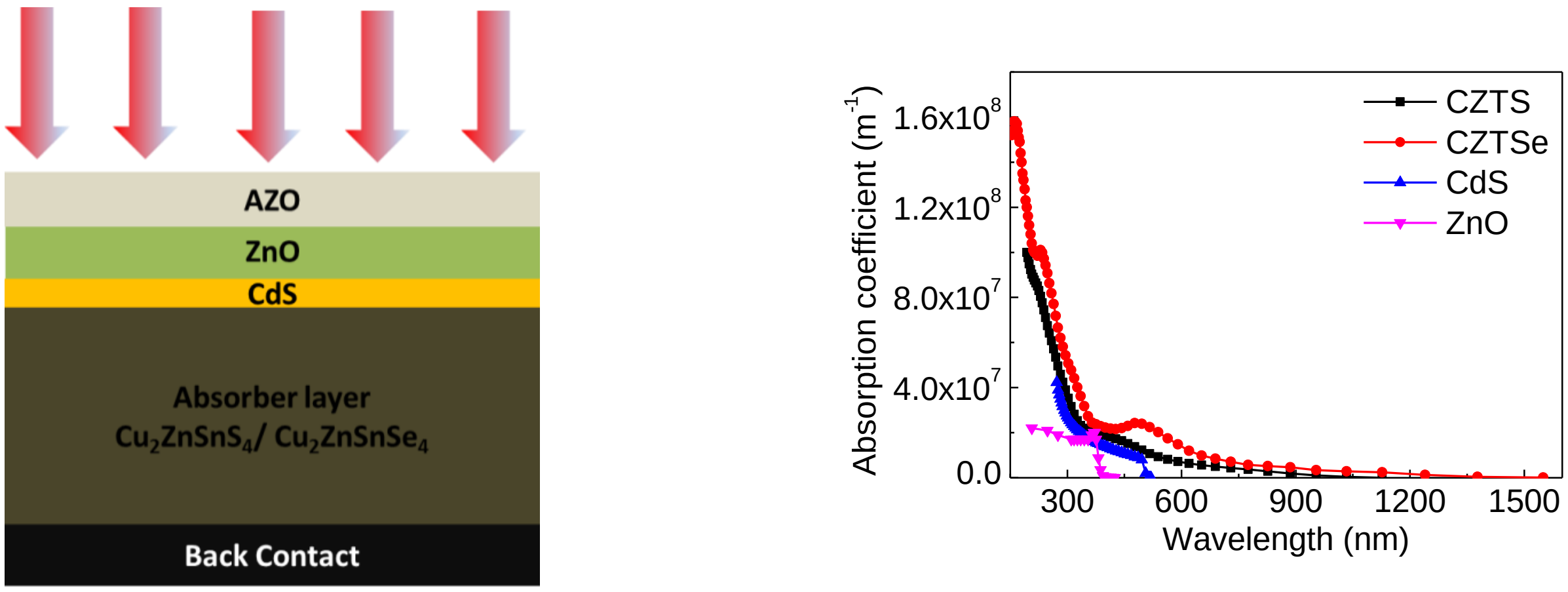

**Figure 1:** (a) Schematic representation of "Al:ZnO/ZnO/CdS/***Absorber layer***/Back contact" single

junction solar cell structure, used for simulating the device performance and (b) the absorption coefficients of different materials used in the present work.

**Table 1:** Material parameters, used in the present work for simulating single and tandem junction solar cell structures

| Material properties | CZTSe | CZTS | CdS | i-ZnO | Al:ZnO |
|---|---|---|---|---|---|
| Thickness [μm] | 2 | 2 | 0.05 | 0.08 | 0.2 |
| Bandgap [eV] | 1 | 1.5 | 2.42 | 3.37 | 3.37 |
| Electron Affinity [eV] | 4.46[13] | 4.3[13] | 4.5 | 4.6 | 4.6 |
| Dielectric permittivity | 9.1[12] | 6.95[12] | 9[14] | 9[14] | 9[16] |
| Density of states in CB [$cm^{-3}$] | $2.2 \times 10^{18}$ | $2.2 \times 10^{18}$ | $1.8 \times 10^{19}$[14] | $2.2 \times 10^{18}$[14] | $2.2 \times 10^{18}$ |
| Density of states in VB [$cm^{-3}$] | $1.8 \times 10^{19}$ | $1.8 \times 10^{19}$ | $2.4 \times 10^{18}$[14] | $1.8 \times 10^{19}$[14] | $1.8 \times 10^{19}$ |
| Thermal velocity of electron [cm/s] | $1 \times 10^{7}$ | $1 \times 10^{7}$ | $1 \times 10^{7}$ | $1 \times 10^{7}$ | $1 \times 10^{7}$ |
| Thermal velocity of hole [cm/s] | $1 \times 10^{7}$ | $1 \times 10^{7}$ | $1 \times 10^{7}$ | $1 \times 10^{7}$ | $1 \times 10^{7}$ |
| Electron mobility [$cm^2/Vs$] | 145 | 100 | 160 | 150 | 150 |
| Hole mobility [$cm^2/Vs$] | 40 | 35 | 50[14] | 25[14] | 25[16] |
| Donor concentration [$cm^{-3}$] | 0 | 0 | $1 \times 10^{17}$ | $1 \times 10^{17}$ | $1 \times 10^{20}$ |
| Acceptor concentration [$cm^{-3}$] | $5 \times 10^{16}$ | $5 \times 10^{16}$ | 0 | 0 | |
| Absorption coefficient [$cm^{-1}eV^{1/2}$] | file[12] | file[12] | SCAPS | SCAPS | |
| Radiative recombination coefficient [$cm^{-3}/s$] | $1.04 \times 10^{-10}$ | $1.04 \times 10^{-10}$ | $1.04 \times 10^{-10}$ | $1.04 \times 10^{-10}$ | |
| Effective mass electron | 0.07 | 0.18[16] | 0.25[14] | 0.275[14] | 0.275[16] |
| Effective mass hole | 0.2 | 0.71[16] | 0.7[14] | 0.59[14] | 0.59[16] |
| Hole capture cross section ($cm^2$) | $1 \times 10^{-15}$ | $1 \times 10^{-15}$ | $1 \times 10^{-13}$ | $1 \times 10^{-15}$ | $1 \times 10^{-15}$[16] |
| Electron capture cross section ($cm^2$) | $1 \times 10^{-15}$ | $1 \times 10^{-15}$ | $1 \times 10^{-15}$ | $1 \times 10^{-15}$ | $1 \times 10^{-15}$ |
| Minority carrier lifetime | 5 ns | 10 ns | | | |
| Interface recombination speed (cm/s) | $10^{3}$[18] | $10^{4}$[18] | | | |
| Defect type at bulk/interface | Donor/Neutral | Donor/Neutral | | | |

The proposed device structures are simulated using the one dimensional numerical simulator SCAPS[19], where coupled Poisson and continuity equations for both electrons and holes with the suitable boundary conditions, defined at the interfaces and contacts are solved numerically[19]. The single junction solar cell structures with CZTS and CZTSe absorber layer are defined in SCAPS, as shown in Fig.2 (a), in conjunction with different materials' thin films stacks and incident light towards the front contact.

This structure, Al:ZnO/ZnO/CdS/***Absorber layer***/Back contact, is considered for the simulation of single junction solar cells with CZTS and CZTSe as the absorber layer materials. Al:ZnO is providing the front electrical contact and simultaneously remains invisible/transparent to the incident solar radiation. ZnO layer serves as the window layer and n-type CdS layer acts as the buffer layer, making a heterostructure p-n junction with p-type CZTS/Se absorber layer.

The donor defects are introduced at 0.6 eV above valance band, in absorber layer and neutral defects at absorber layer/CdS hetero-interface with energy 0.6 eV above valance band. The defect energy level are taken to ascertain mid gap defect. Mo back contact is considered with work function about 5 eV and surface recombination speed ~ $1 \times 10^{5}$ cm $s^{-1}$ and $1 \times 10^{7}$ cm $s^{-1}$ for electrons and holes, respectively at the back contact. Thermionic emission is considered for the transport properties of majority charge carrier. The band to band recombination is considered for the bulk materials and also for radiative recombination with recombination coefficient equal to $1.04 \times 10^{-10}$ $cm^{3}s^{-1}$ for all the layers in the simulation[20]. Carrier transport through tunneling mechanism is also considered at each interfaces and the effective mass considered during the simulation is given in Table 1. Auger recombination is not considered for the present studies, as it is significant only at higher carrier concentrations. For simplicity of calculation capture cross section and thermal velocity for electron and holes are kept same $10^{-15}$ $cm^{2}$ and $10^{7}$ cm/s respectively. The minority carrier life time at the bulk of absorber layer and the interface recombination speed at absorber/buffer layer interface are taken 10 ns, 5 ns, and $10^{4}$ cm $s^{-1}$ ,$10^{3}$ cm $s^{-1}$ for CZTS and CZTSe respectively [18]. This assumption for the base device falls well within the range of mostly reported kesterite devices. The absorption coefficient as a function of wavelength is shown in Fig 1(b) for different layer of the devices, used in the simulation. This is more realistic in contrast to considering a constant value for the simulation, showing more insights in the device performance.

**Device analysis for a single junction solar cell with CZTS and CZTSe absorber layers:**

The single junction solar cell device structures with CZTS and CZTSe absorber layers, as shown in Fig. 1(a), are first simulated using various materials parameters, as summarized in Table 1 for both CZTS and CZTSe absorber layers. The estimated current voltage characteristics are shown in Fig. 2(b) for devices based on these absorbers, with respective energy band diagrams, Fig. 2(a), showing the alignment of respective bands near interfaces for these solar cell structures. We observed spike like band alignment at CZTS/CdS interface. The band alignment at the heterointerface is critical for the device performance. The large cliff in the band structure at heterointerface lowers the open circuit voltage and spike type heterostructure reduces the short circuit current density (Jsc) of the device. That's why the optimization of heterostructure interface is necessary for enhanced performance. The photovoltaic characteristics for single junction device showed large short circuit current density for CZTSe absorber and large open circuit voltage for CZTS absorber based devices, as shown in Fig. 2(b). The observed difference in short circuit current density can be well understood by observing the quantum efficiency plot shown in the inset of Fig. 2(b). The quantum efficiency curve covers large area owing to the smaller band gap of CZTSe absorber layer, thus allowing lower energy photons to participate in generating photocurrent, thereby increasing the short circuit current density of the cell. High open circuit voltage for the CZTS absorber cell is due to its larger band gap as compared to that of CZTSe absorber. The optimal photovoltaic efficiencies observed are ~ 13.41% and 14.88% for CZTS and CZTSe absorber based photovoltaic devices, respectively. The respective device parameters are summarized in a table as an inset in Fig 3(b).

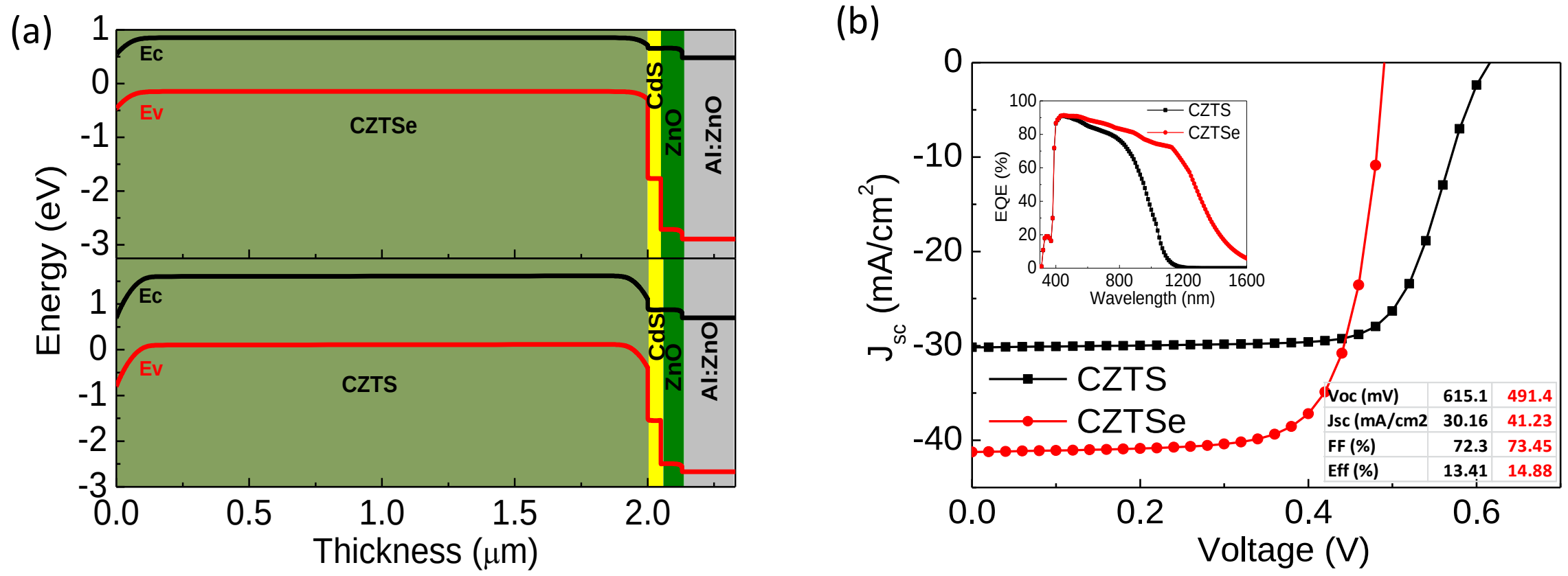

**Figure 2:** (a) Band alignments at different interfaces and (b) Current-voltage (I-V) characteristics for single junction CZTS and CZTSe cells with insets showing external quantum efficiency (EQE) versus wavelength variation (top left) and device parameters (bottom right table).

**Impact of absorber layer thickness on single junction photovoltaic performance:**

The influence of absorber layer thickness is investigated for these device structures and the observations are summarized in Fig. 3. We found that CZTS solar cells exhibit relatively larger open circuit voltage, Fig. 3(a), as compared to that of CZTSe absorber layer based solar cell. However, there is no significant effect of thickness on open circuit voltage (Voc) of the device and a small increase with thickness in noticed, approaching towards the optimal Voc for CZTS. This increase is attributed to the overall increase in majority carriers in bulk absorber, as detailed in next section. The short circuit current density (Jsc) is shown in Fig. 3(b) for these cells, increases significantly with increasing the thickness of absorber layers. The cumulative effect results in the increased device efficiency, Fig. 3(c). The efficiency of these single junction solar cells increases with increasing thickness and showing saturation after a critical thickness. The thickness of about 2 μm seems sufficient to achieve the maximum photovoltaic response. This result is in agreement with Beer-Lamberts law $I = I_0 e^{-\alpha x}$, suggesting enhanced absorption in the absorber layer resulting into the excess photogenerated carriers and thus, improved device efficiency. The variation in quantum efficiency against wavelength is shown Fig. 3(d) for CZTS absorber solar cell, with inset showing the same for CZTSe solar cell for different absorber thicknesses. A large area is covered under the quantum efficiency curve for CZTSe absorber cell as compared to that of CZTS absorber cell because of its lower band gap. The lower band gap of CZTSe also allows the enhanced absorption at lower wavelength, resulting in improved quantum efficiency. This is also reflected in the current density plot, Fig. 3(b), showing relatively larger current density for the CZTSe cell as compared to that of CZTS cell for same thickness.

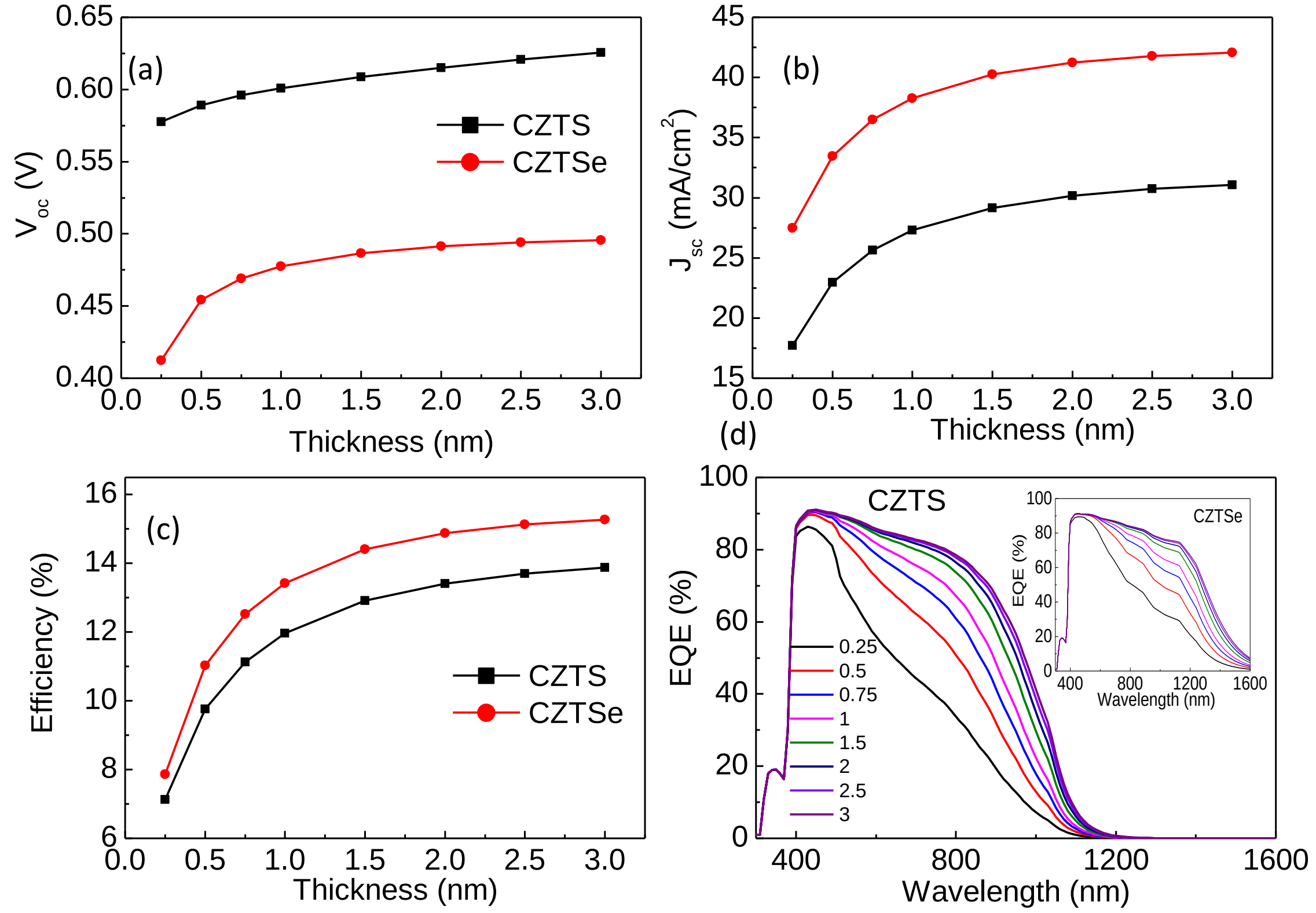

**Figure 3:** (a) Open circuit voltage; (b) short circuit current density; (c) efficiency, and (d) quantum efficiency of CZTS and CZTSe absorber solar cells for different thickness of the absorber layer.

**Impact of absorber layer acceptor concentration on single junction photovoltaic performance**

CZT(S/Se) is an intrinsic p-type photovoltaic absorber material, where carrier concentration relies on the non-stoichiometry and defects present in the synthesized material. The carrier concentration is varied from 1 x$10^{12}$ $cm^{-3}$ to 1 x$10^{18}$ $cm^{-3}$ for the absorber layer in the present work. The open circuit voltage, $V_{oc,}$ Fig. 4(a), and the short circuit current density, $J_{sc}$ , Fig. 4(b), increases steadily and then starts decreasing, with increasing the carrier concentration. The calculated results are in agreement with the fact that increased carrier concentration should reduce the minority carrier life time, implying more recombination and thus, reducing the collection of charge carriers at the contacts. Further, increased carrier concentration decrease the depletion width towards absorber layer, thus affecting the effective separation of photo generated charge carriers and causing the reduction in the short circuit current density. The open circuit voltage increases with doping initially and then decreases, because of the compensating nature of donor type defects considered for the system and thus, optimum carrier concentration should be chosen in such a way that the large peak power can be achieved. The simulated efficiency versus acceptor density results (see Fig 4(c)) suggest that the optimal carrier density is up to 5 x $10^{16}$ $cm^{-3}$ for CZTS and 1 x $10^{17}$ $cm^{-3}$ for CZTSe solar cell to achieve the maximum photovoltaic performance. The calculated maximum efficiency, $V_{oc}$ and $J_{sc}$ with these carrier densities are ~13.41%, 615m V and 30.15 mA/$cm^2$, respectively for CZTS and ~15.6%,   514.29 mV, 40.52 mA/$cm^2$ for CZTSe solar cell, respectively.

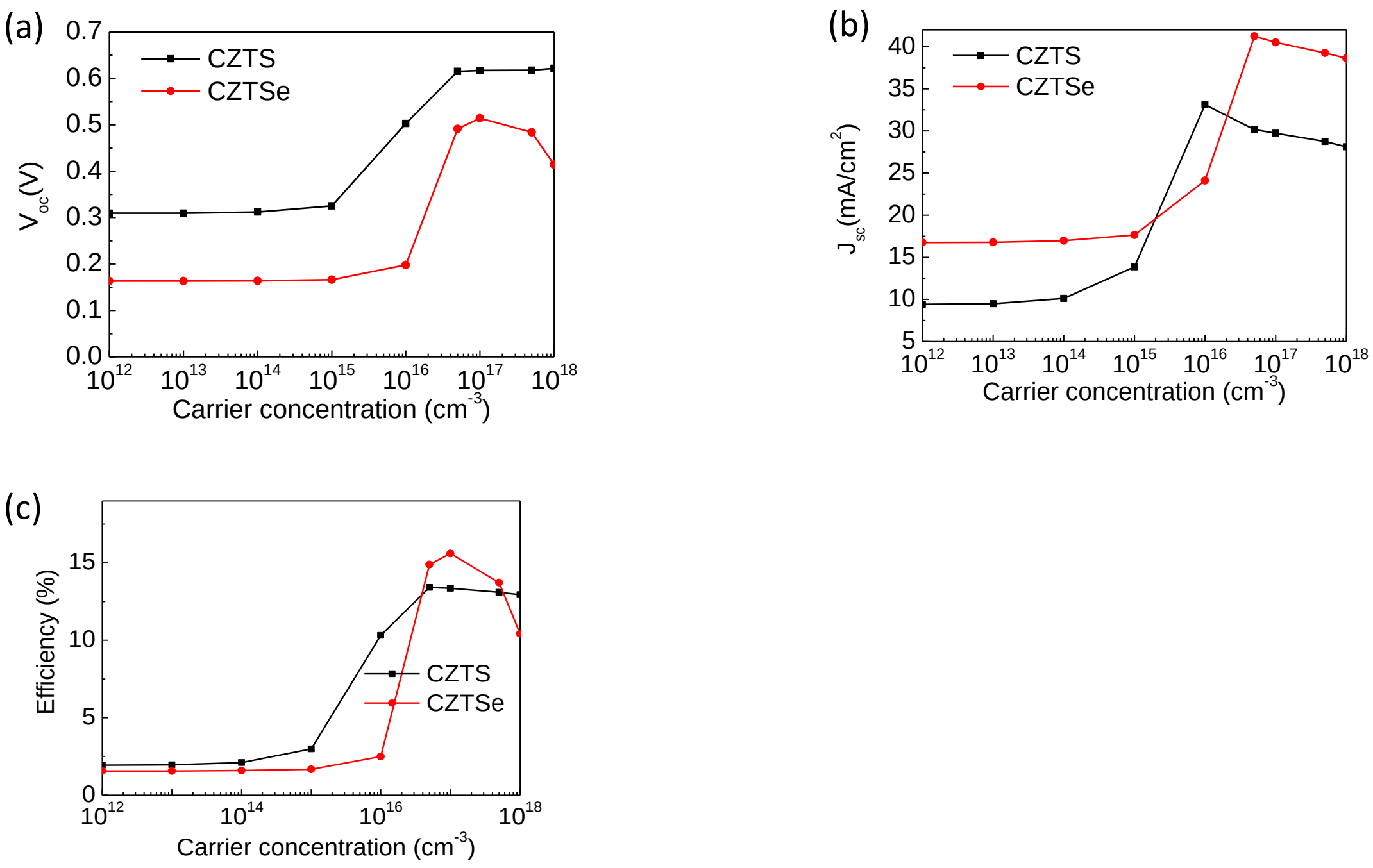

**Figure 4:** (a) Open circuit voltage; (b) short circuit current density; and (c) efficiency as a function of carrier concentration for CZTS and CZTSe absorber based solar cells.

**Impact of defect concentration at absorber and at absorber/CdS interface**

CZTS,Se is a defect prone system and consists of acceptor and donor vacancies, antisites and interstitials. Low formation energy of acceptor defects in the CZTS,Se system makes it an intrinsic p-type system without additional foreign dopant elements. There is provision to introduce different donor and acceptor defects in SCAPS, however available characteristic data for all kinds of defects possible in CZTS,Se system is rare. For simplicity of the simulation, we considered single donor bulk defects in the absorber layer and neutral defects in absorber/CdS interface. The defects are considered at mid band with characteristic energy level at 0.6 eV above valance band. The concentration of bulk defects determines the lifetime of minority carrier in the absorber while defects at the interface determines the recombination speed (S) at the interface varying the defect density in the absorber and absorber/CdS interface. The change in efficiency is summarized in Fig 5(a) and 5(b) against absorber and absorber/CdS interface defect concentrations, respectively. We can see that for moderate defect concentrations, especially in absorber, the efficiency is not changing significantly. This is probably due to the lower defect-photogenerated carrier scatterings because of lower defect concentration and thus, not affecting the device performance. However, after certain defect concentration, in the range of high $10^{16}$ $cm^{-3}$, the efficiency drops drastically to the very low values, close to zero for both CZTS and CZTSe absorbers. In contrast, absorber/CdS interface defects seems more prone in degrading the photovoltaic performance and any increase in interface defect concentration resulted in reduced photovoltaic efficiency, Fig 5(b). Thus, absorber/buffer interface defects should be minimized to a possible lower level to realize the enhanced photovoltaic efficiency. Further, we investigated the variation in absorber/CdS interface recombination (IR) speed against minority carrier life time for different photovoltaic efficiencies and results are summarized as contour plots in Fig 5(c) and 5(d) for CZTS/CdS and CZTSe/CdS interfaces, respectively. These measurements suggest that higher minority carrier life time and lower interface recombination (IR) speeds are very critical to realize the enhanced photovoltaic response for both CZTS and CZTSe absorber based photovoltaic cells.

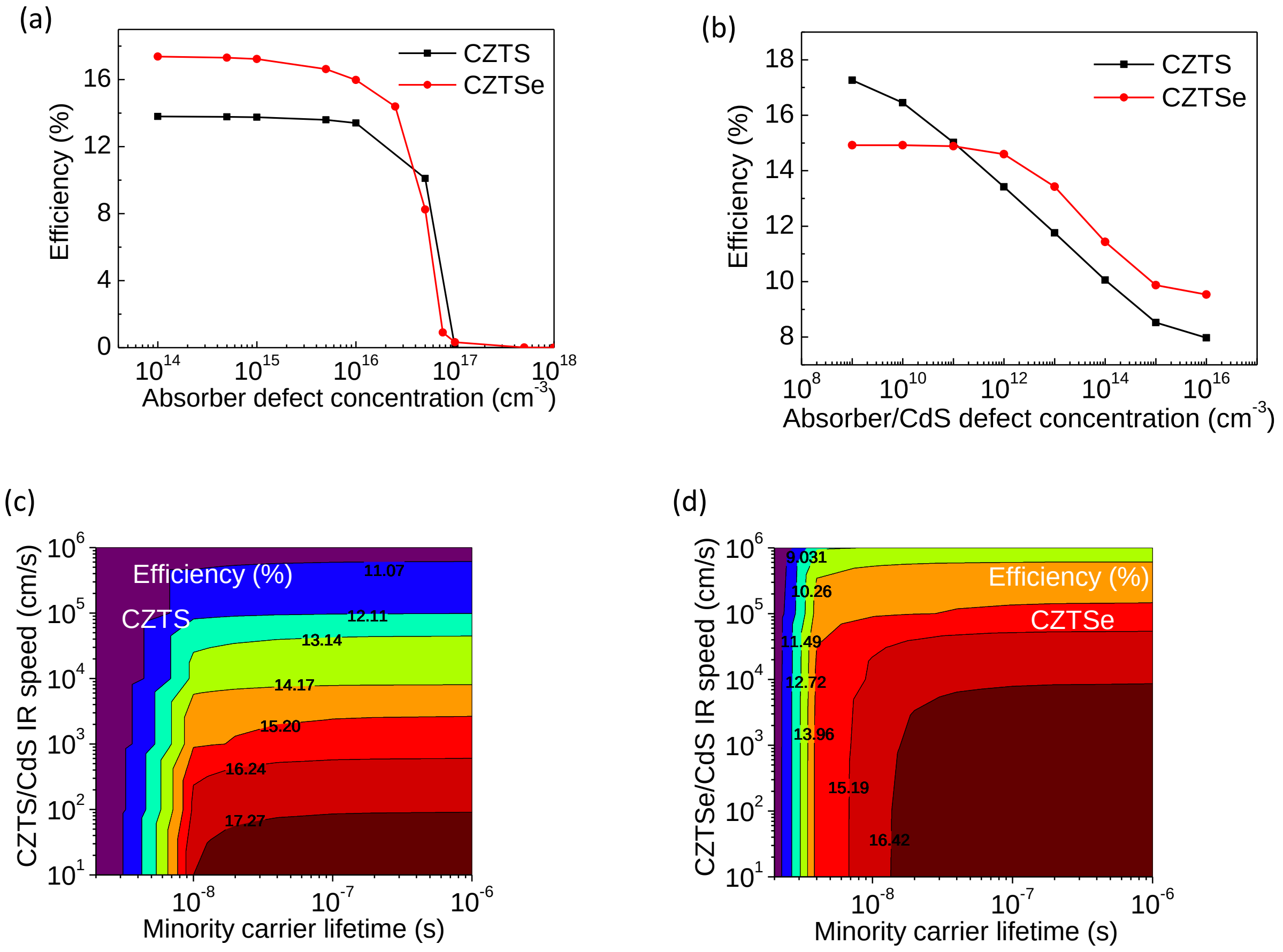

**Figure 5:** Impact of defect concentration (a) and absorber/CdS interface defect concentration (b) in efficiency for CZTS and CZTSe solar cells and contour plots showing variation of efficiency with minority carrier life time and interface recombination speed in CZTS (c); and CZTSe (d).

## Impact of buffer layer

CdS is the most commonly used buffer layer in CZTSSe solar cell. The performance of a solar cell depends on the band alignment at the heterointerface. CdS with wide bandgap of ~2.42 eV allows the maximum photoabsorption in the absorber layer. The thickness of a buffer layer mainly depends on its conductivity. Higher is the conductivity of buffer layer, more will be the penetration of depletion region towards the absorber side and thus the higher device efficiency. It is suggested that a very thin layer of CdS is required for enhanced solar performance. The thin layer of CdS will also assist in avoiding absorption loss in CdS buffer layer. Considering the same, thickness of CdS is varied at different carrier concentration and the variation of carrier concentration as a function of CdS buffer layer thickness is plotted as contour plots for different efficiency in Fig 6(a, b) for both CZTS and CZTSe absorber based solar cell devices. These results suggest that the higher CdS carrier concentration is essential for the optimal device performance. In contrast, the device efficiency is relatively insensitive to its thickness for CZTS, while for CZTSe at lower CdS carrier concentration about 55nm of thickness is required to achieve the constant efficiency, Fig 6(b). Further, increase in carrier concentration of CdS buffer layer does not show any significant improvement on the device performance.

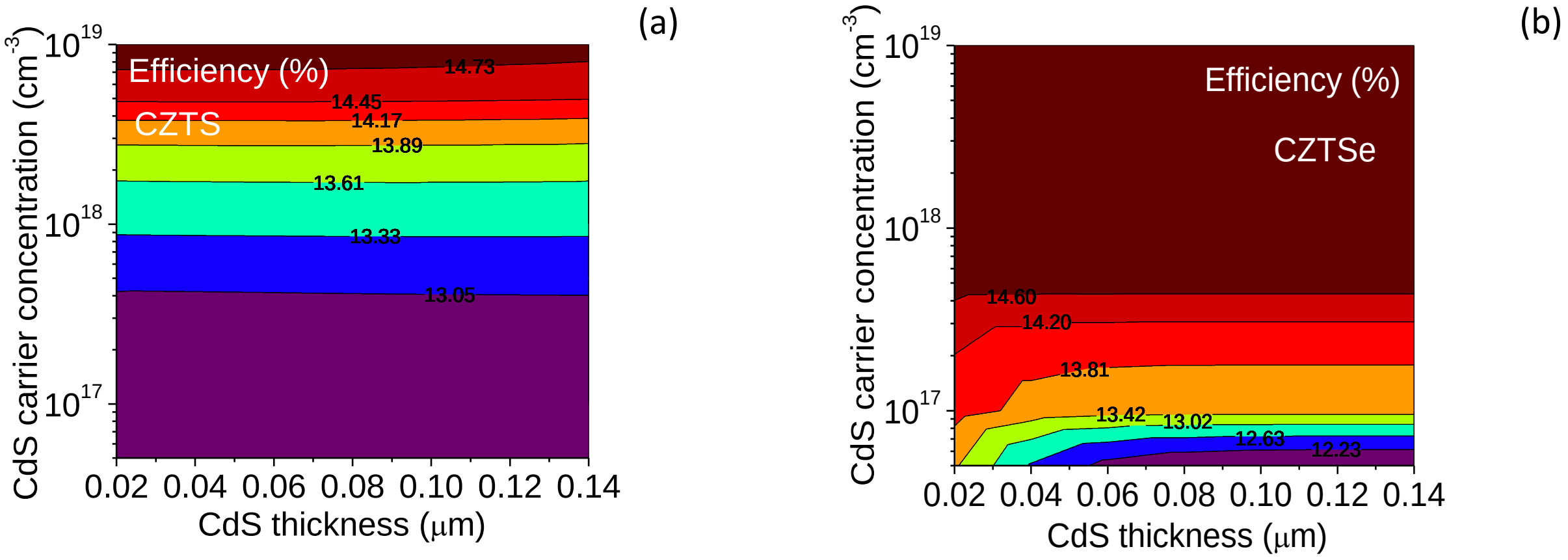

**Figure 6:** Impact of CdS buffer layer thickess and carrier concentration on CZTS (a) and CZTSe (b) solar cell efficiency

The conduction band offset (CBO) at the absorber/buffer interface plays an important role in governing the carrier transport through the junction. The offset is primarily determined by the difference in the values of electron affinity of the absorber and buffer layers. The positive and negative band offsets govern the spike and cliff like heterostructures, respectively. In present simulation, electron affinity of CdS is varied to see the effect of band offset at hetero interface in the device performance. Conduction band alignment at the heterostructure interface is shown in Fig 7(a, b) for CZTS and CZTSe solar cells. As we go on increasing the electron affinity of buffer layer the band alignment at the absorber/buffer layer interface changes from spike to cliff.

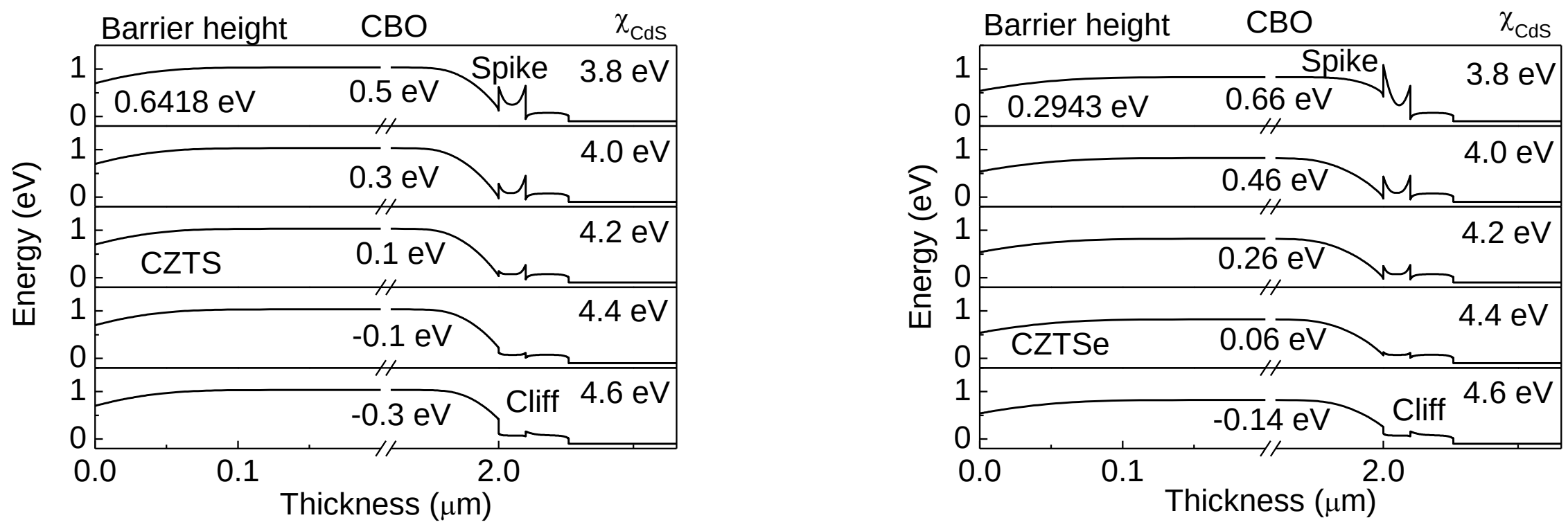

**Figure 7:** Conduction band offset (CBO) in (a) CZTS (b) CZTSe at different CdS layer electron affinity ($\chi_{CdS}$)

We observed that a cliff in the heterostructure results in a decrease in open circuit voltage, Fig 8; whereas short circuit current remains relatively unchanged, except at very low CdS electron affinity values. Further, the efficiency versus CdS electron affinity variation is summarized in Fig 8 (c), showing inverted U shaped behavior. Thus, a moderate electron affinity range 4.0 eV – 4.4 eV is essential to realize the maximum efficiency for both CZTS and CZTSe absorber based photovoltaic devices.

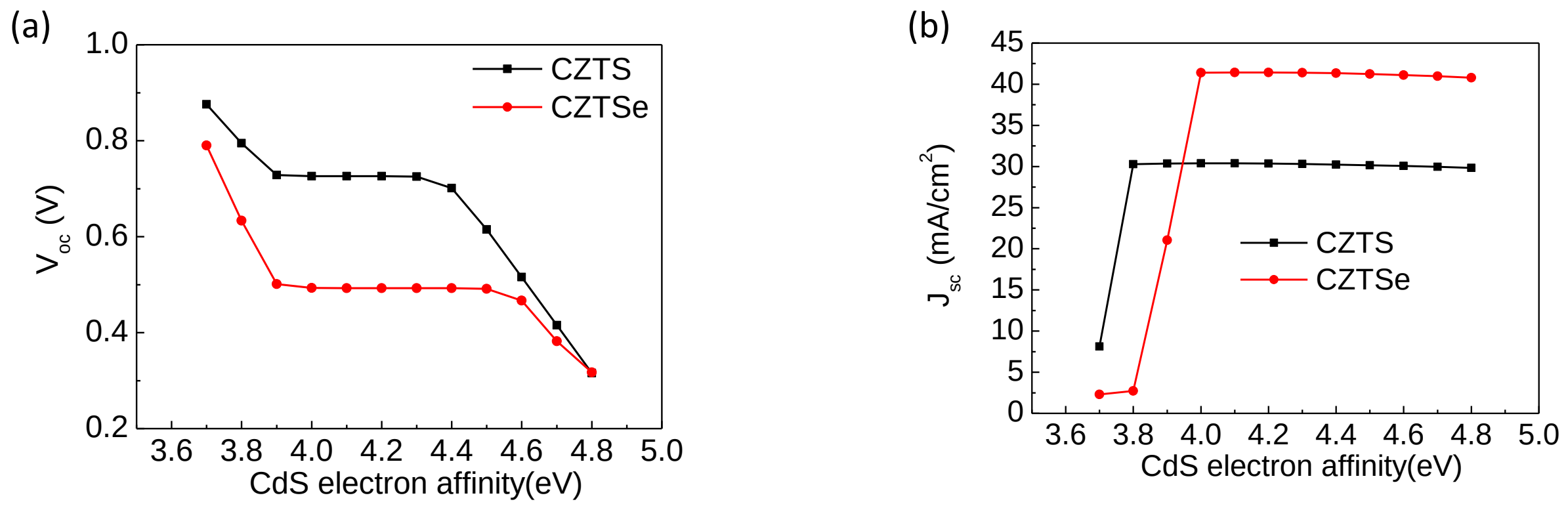

(c)

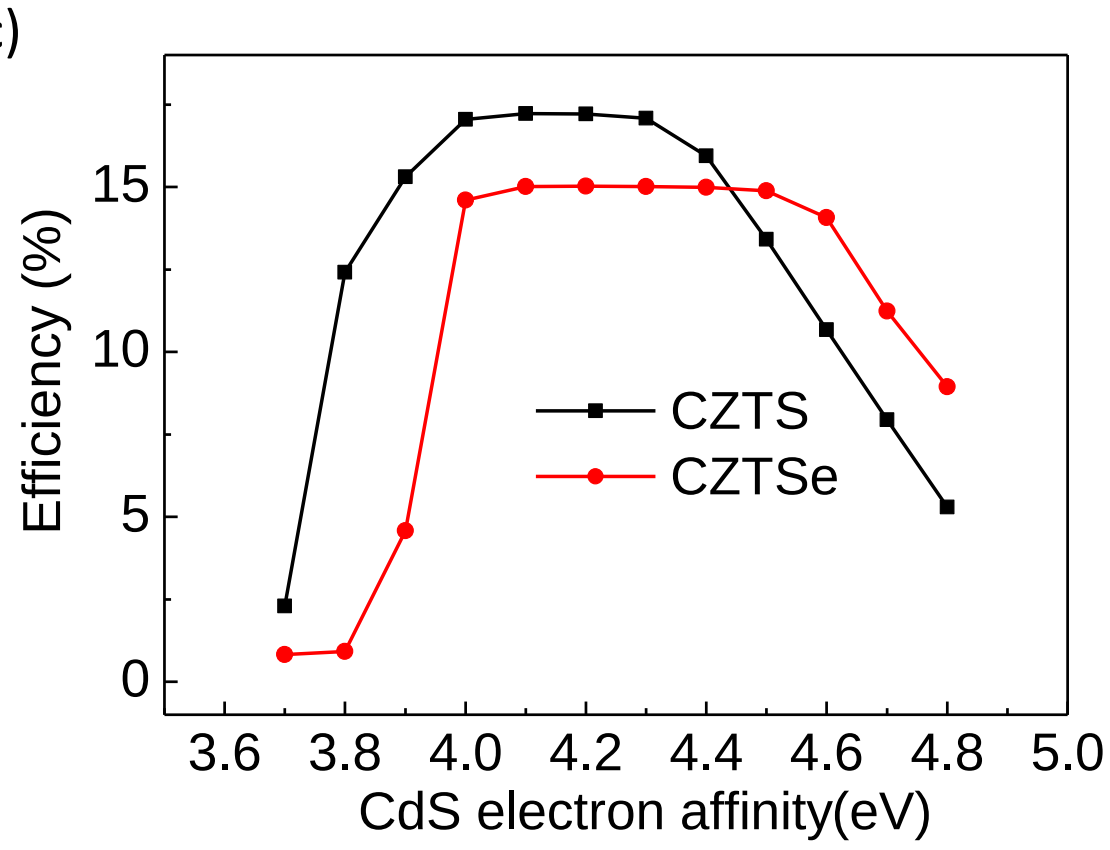

**Figure 8:** Impact of CdS electron affinity (band offset at absorber/CdS interface) on Voc (a); Jsc (b); and efficiency (c) for CZTS and CZTSe solar cells

These observations suggest that the choice of buffer layer material with CZTSSe absorber material is crucial and should provide the desired band offset for synthesized photovoltaic devices. We further carried out studies for different electron affinities of the buffer layer material and the interface recombination (**IR**) speed on open circuit voltage, short circuit current and efficiency, to explore the selection parameters/criteria for a buffer. The results are summarized in Fig 9 as contour plots between absorber layer/CdS interface recombination (IR) speed and CdS electron affinity with open circuit voltage, short circuit current and efficiency. It can be seen from the contour plot of $V_{oc}$ Fig 9 (a & b) and $J_{sc}$, Fig 9 (c & d) that cliff in the band structure ($\chi_{CZTS,Se} < \chi_{Buffer}$) affects Voc of the device, whereas Jsc is nearly insensitive to the same for both CZTS and CZTSe absorber based photovoltaic devices. $V_{oc}$ decreases with the increase in cliff at the heterointerface. In contrast, $J_{sc}$ showed decrease with spike ($\chi_{CZTS,Se} > \chi_{Buffer}$) like band offset at the heterointerface. The interface recombination (IR) speed is showing a negative impact on the performance and thus, a low interface recombination speed is desired for enhanced photovoltaic response.

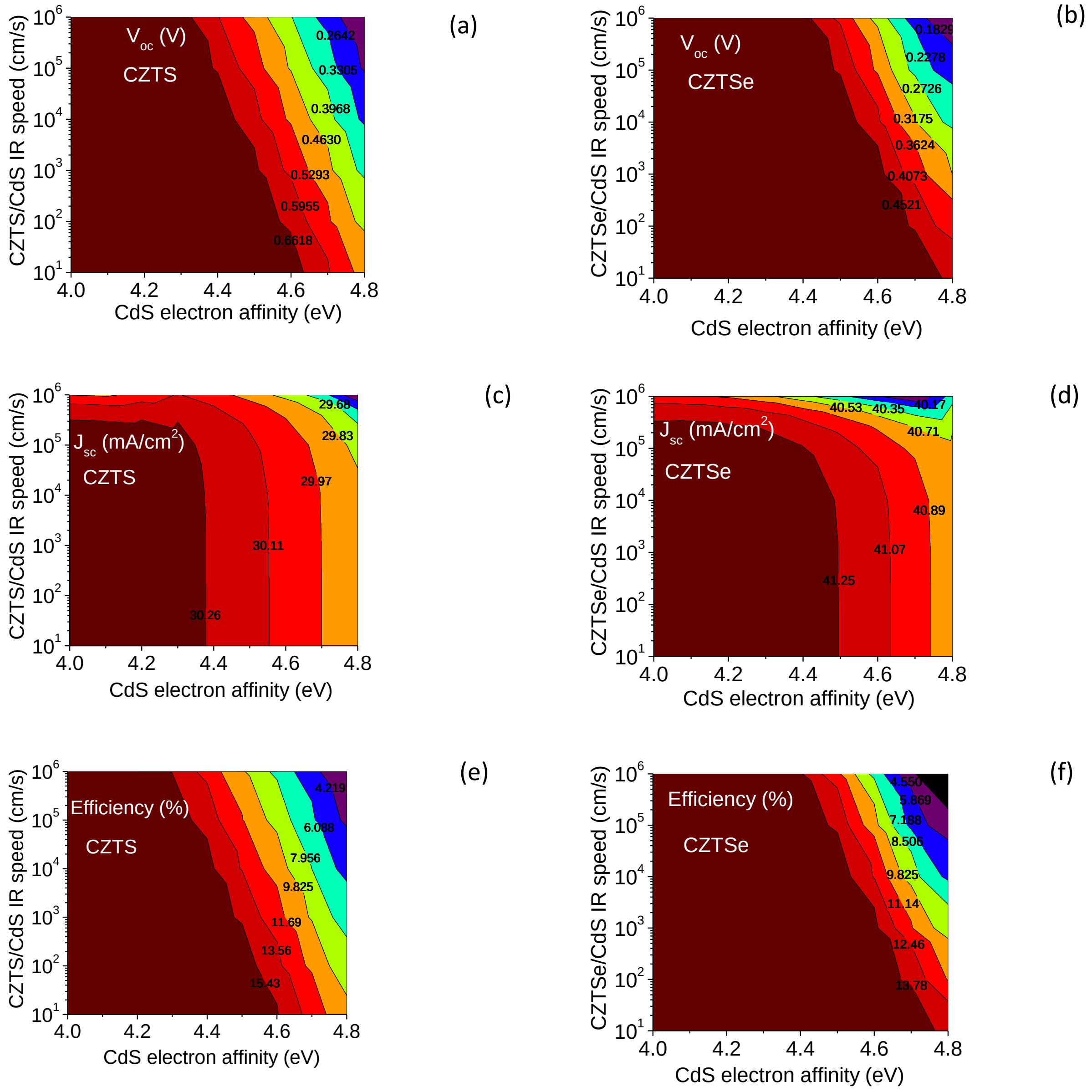

**Figure 9:** Impact of interface recombination speed and CdS electron affinity on CZTS (a) Voc,(c) Jsc (e) efficiency and CZTSe (b)Voc (d) Jsc (f) efficiency

### Impact of Back contact work function

The back contact material shows severe impact on the solar cell performance. A relatively higher metal work function is required to realize an ohmic contact to the absorber material. Molybdenum is commonly used as back contact material for CZTSSe absorber based photovoltaic devices. However, molybdenum is not the optimum choice as the back contact because of its non-ohmic contact characteristics with kesterite

CZTSSe absorbers. We systematically investigated the impact of back contact work function by varying it from 4.4 to 5.5 eV and the results are summarized in Fig 10. Increasing metal work function decreases the barrier height for majority charge carrier at back contact interface thereby improved performance is observed. The open circuit voltage, short circuit current and efficiency show the similar trend, increasing initially with increasing metal work function and above a critical metal work function, these values show saturation behavior. Thus, a high metal work function is required for the enhanced solar cell performance, which will allow Ohmic-contact formation at the back electrode. These observations suggest that metal work function above 5.3 eV and above 5 eV is required for CZTS and CZTSe absorber based solar cell.

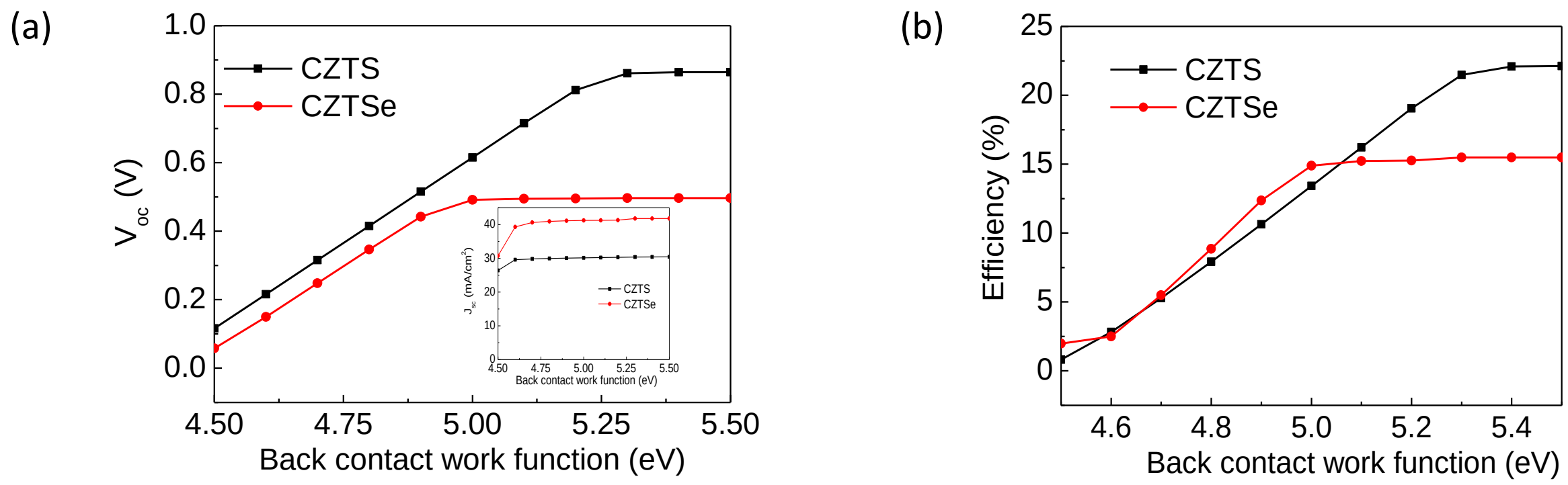

**Figure 10:** Impact of metal work function on Voc (inset Jsc) (a) and efficiency (b) for CZTS and CZTSe solar cell

**Design and analysis of CZTS/CZTSe tandem solar cell structure photovoltaic device:**

Single junction solar cell performance is limited by the photon absorption for the respective band gap of absorber layer of the cell. The performance of the solar cell can be improved by stacking single junction solar cell with different band gap absorbers in such a way that the higher energy photons are absorbed in the top cell and the lower energy photons are absorbed in respective lower solar cells. In CZTS/CZTSe tandem cell, as shown in Fig 11, top cell is made of larger band gap CZTS absorber layer (Eg ~ 1.5 eV) and the bottom cell is made of lower band gap CZTSe bottom cell (Eg ~ 1 eV). The tandem cell structures can be realized by fabricating the top cell on the transparent conducting surface on the upside layer of the bottom cell. For example, S. Nishiwaki et al. fabricated the stacked chalcopyrite tandem solar cell by connecting $CuGaSe_2$ and $CuInGaSe_2$ cells in series and reported the cell efficiency ~ 7.4% with ~ 1.2V open circuit voltage [10].

The fabrication of tandem cell structures is possible using monolithic integration of one cell on the top of other cell. The stacking of such two cells forms a reverse biased p-n junction at the interface that hinders the current flow across the device. An n+/p+ tunnel junction is essential at the interface to overcome the reverse bias issues and for proper current flow across the tandem structure. The transparent materials such as indium tin oxide (ITO), fluorine doped tin oxide (FTO), aluminum doped zinc oxide (AZO), molybdenum doped indium oxide (IMO) with high mobility can serve well in making tunnel junction between the cells. Tokio Nakada et al. fabricated chalcopyrite tandem solar cell with $Ag(In_{0.2}Ga_{0.8})Se_2$ (AIGS) upper cell on top of the transparent IMO and CIGS as the bottom cell. This tandem structure showed ~ 8% photovoltaic efficiency with 1.3V open circuit voltage [11].

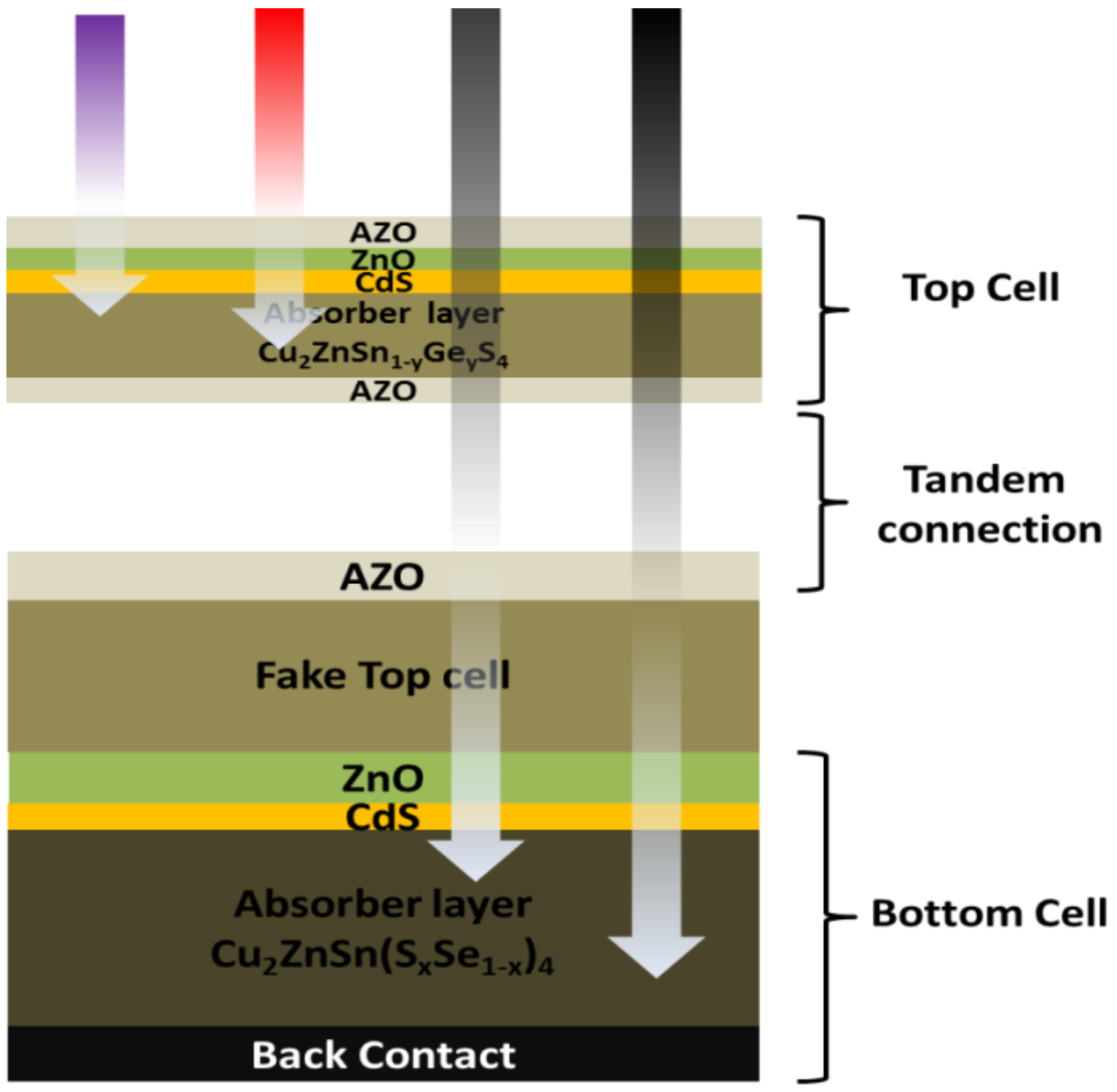

**Figure 11:** Schematic representation of tandem cell structure with CZTS top cell and CZTSe bottom cell in conjunction with tandem connections

The simulation of tandem structure is quite cumbersome in SCAPS and needs to be considered carefully by utilizing the concept of replacing the actual cell by an equivalent fake absorber layer. This can be simulated by connecting the top and bottom cell in series, which are simulated independently for optimal performance. However, the problem persists as the bottom cell should be illuminated through the adapted solar spectrum, the residual solar spectra after its partial absorption across the top cell. A top cell is considered as a fake CZTSe layer of same thickness as the original bottom cell to overcome this problem. The fake CZTSe layer is considered p+ layer with its electron affinity to avoid any band discontinuity and to ensure the optical presence and electrical absence of this fake CZTSe layer, while simulating the top cell. Additionally, high defect density is considered for fake CZTSe layer to reduce the minority carrier diffusion length and to ensure that lower energy spectrum (hν < 1.5 eV) is not contributing to the photocurrent of the tandem structure.

A fake CZTS layer is considered which can absorb the substantial fraction of the incident light as in the real tandem cell to set the bottom cell for the tandems structure. This CZTS layer is taken as n+ layer with sufficient electron affinity to avoid any band discontinuity for electrically inactiveness of this layer. The high defect density is taken in the fake CZTS layer for small minority carrier diffusion length and to ensure high energy solar spectra (hν > 1eV) is not contributing to the photocurrent of the device.

The maximum current of a tandem cell is limited by the minimum current produced by the constituent solar cells (sub cell) and the maximum open circuit voltage is limited by the sum of open circuit voltage of the individual solar cells. Thus, the individual constituent cells do not limit the current conduction to get the maximum advantage of the tandem cell. The constituent cells should be made of different bandgap absorbing materials, which may cover the maximum solar spectrum while current is matched across these cells. This can be achieved from the current density thickness profile for the tandem structure as shown in Fig 12(a). The current matched conditions for the tandem structure are obtained at the point of intersection of top and bottom cell current densities, see Fig 12(a). The matched short circuit current density values are summarized in Fig 12(b) for top and bottom cell absorber thickness. We observed that the thickness of the top CZTS cell for matched tandem cell structure varies significantly for different bottom CZTSe cell thickness, Fig 12(b). This variation suggests that the current matching condition can be achieved at relatively lower CZTS top cell thickness as compared to that of the bottom CZTSe cell thickness, Fig 12(b). The tandem cell response improves with increasing the bottom CZTSe cell thickness, as seen in the enhanced tandem cell current density, Fig. 12(a & c).

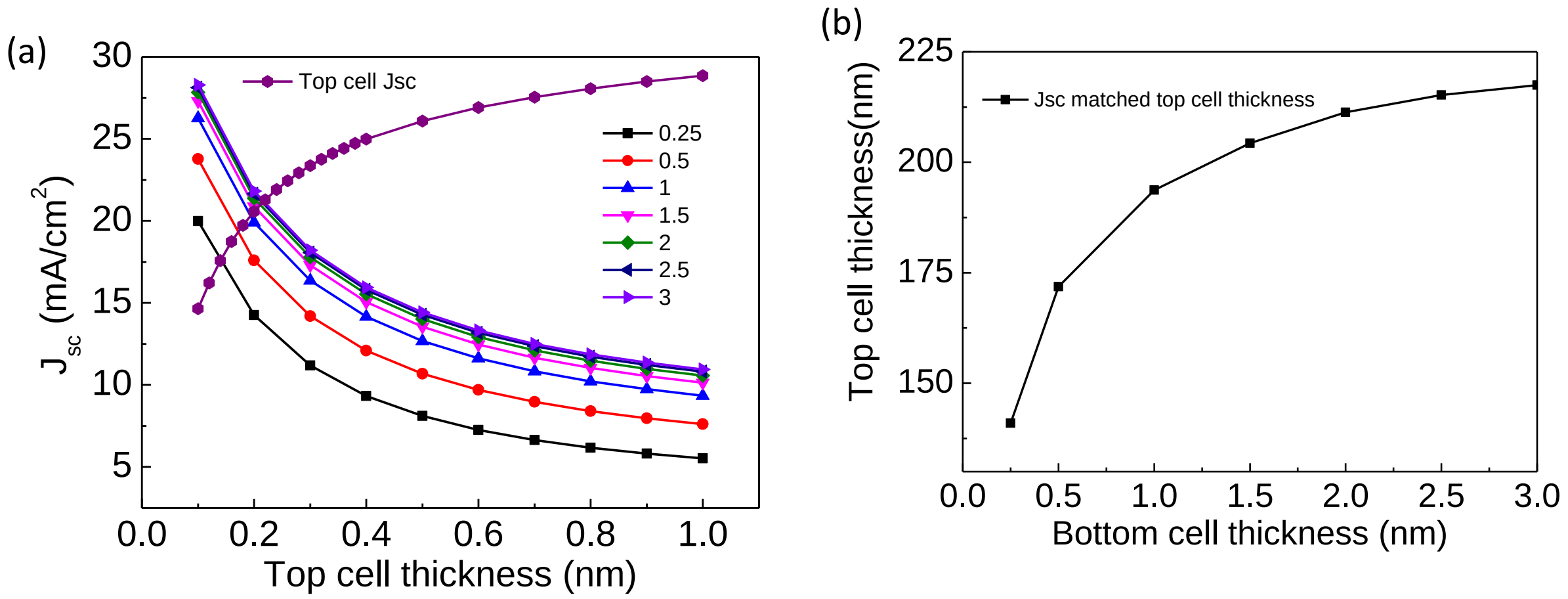

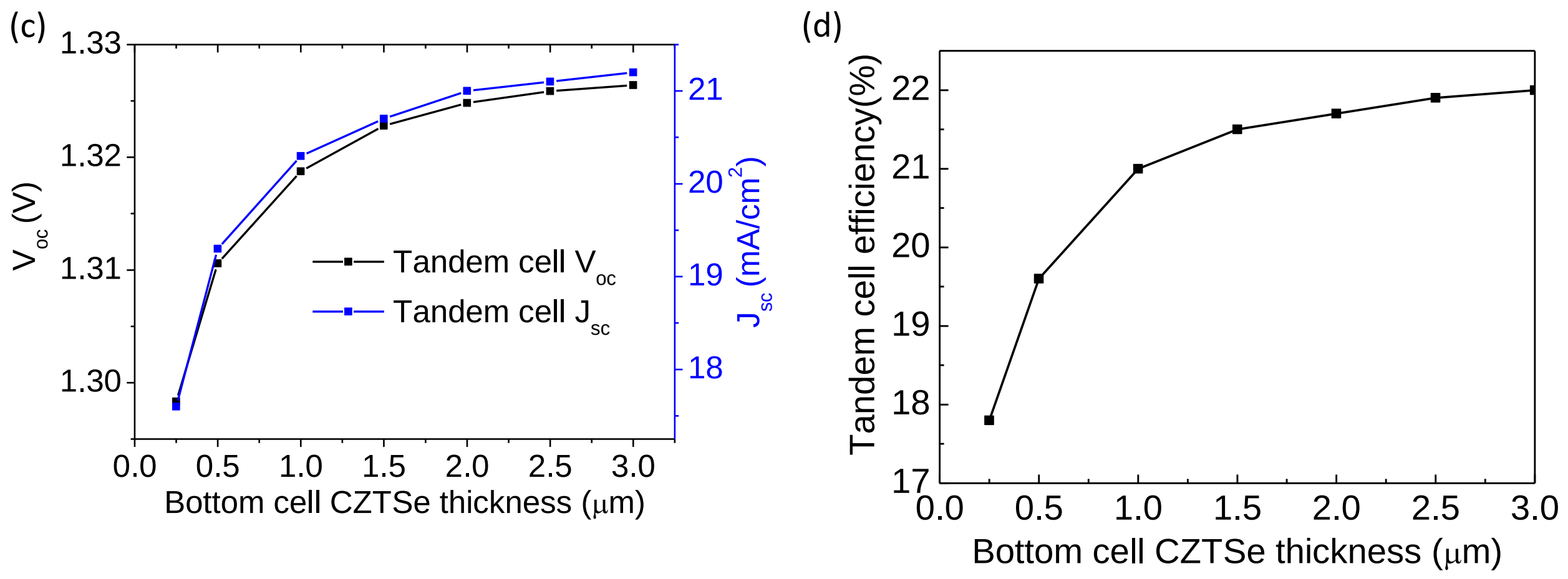

**Figure 12:** (a) Variation of current density of CZTS top cell and CZTSe bottom cell at different thicknesses (b) Thickness of top CZTS cell matched with different thickness of bottom CZTSe cell (c) Current density and open circuit voltage curve for tandem structure with current matched top and bottom cell at different bottom cell thickness (d) Tandem structure efficiency at different bottom cell thickness.

The variation in tandem cell current density and open circuit voltage are shown in Fig. 12(c) for different bottom cell thickness. The increased current density is attributed to the enhanced absorption of solar energy resulting into the enhanced photogenerated charge carriers, and finally the enhanced open circuit voltage. The tandem cell efficiency as a function of bottom cell thickness is shown in Fig. 12(d) under current matching condition. We observed that tandem cell efficiency increases with cell thickness and saturates at or above 2μm thick bottom cell. The current - voltage characteristics and the quantum efficiency are shown in Fig. 13 (a & b) for the tandem structure at 2μm CZTSe thick bottom cell in conjunction with 211 nm thick matched CZTS top cells. We observed the enhanced absorption of photon energy at longer wavelength from a lower bandgap bottom cell and at shorter wavelength from a higher bandgap top cell, resulting in photo current generation over a wide spectral range as seen in the quantum efficiency curve Fig 13 (b). The advantage of the tandem structure can be seen in terms of improved open circuit voltage, which is the sum of open circuit voltages of top and bottom cells. However, the maximum current is limited by the current from CZTS top cell in the tandem cell configuration. This assisted in achieving the higher open circuit voltage for the complete tandem structure with optimal current density. The quantum efficiency (QE) curves are shown in Fig 13(b) for the top CZTS and bottom CZTSe cells. These QE versus wavelength plots indicate that up to 826 nm ($h\nu > 1.5$ eV) top cell is dominating, whereas for longer wavelength ($h\nu < 1.5$ eV) top cell remains invisible and significant absorption takes place in the bottom cell, and thus improving the overall quantum efficiency for tandem structure in wide spectral range. The estimated device parameters are listed in Table 2 for top, bottom and tandem cell structures. The maximum efficiency about 21.7% is achieved for the optimized tandem CZTS/CZTSe solar cell structure in the investigated device configuration.

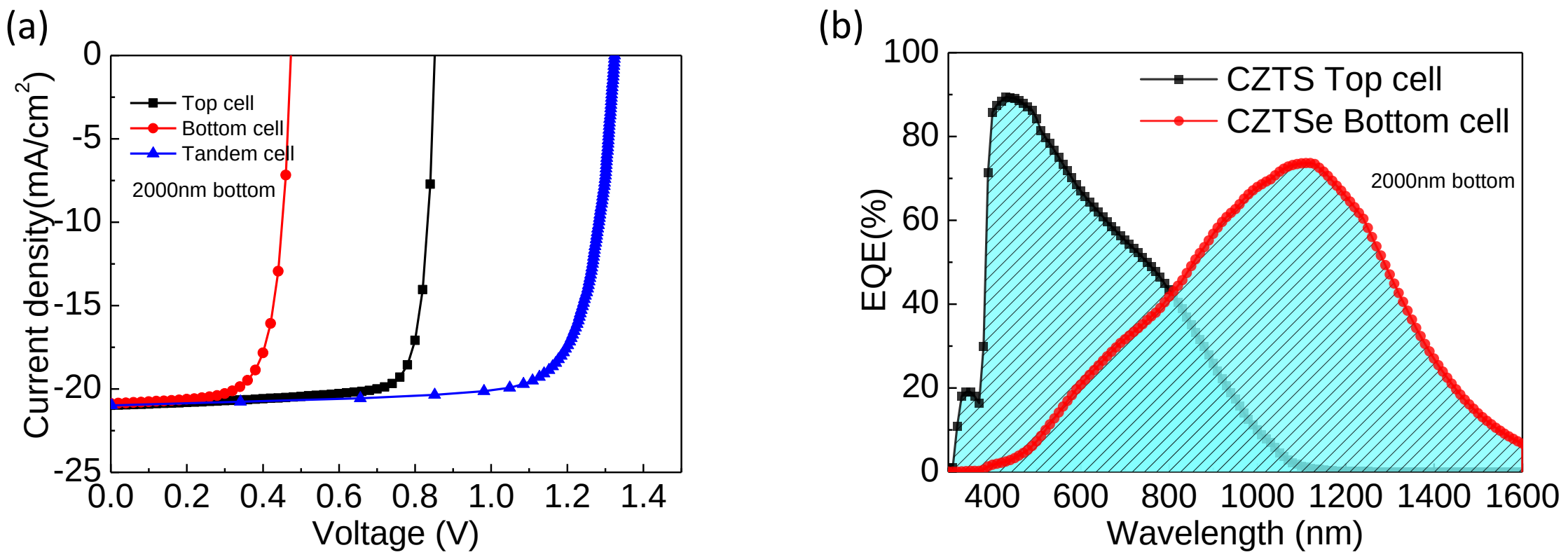

**Figure 13:** (a) Current-voltage characteristics and (b) quantum efficiency for CZTS top cell CZTSe bottom cell and CZTS/CZTSe tandem cell

**Table 2:** CZTS and CZTSe absorber based photovoltaic cell performance parameters for single junction and tandem cell configurations

| Photovoltaic cell configuration with absorber thickness | Jsc ($mA/cm^2$) | $V_{oc}$ (mV) | FF (%) | Efficiency (%) |
|---|---|---|---|---|
| 211.33nm CZTS single cell | 20.98 | 852.9 | 81.95 | 14.67 |
| 2 μm CZTSe single cell | 41.73 | 496.6 | 74.61 | 15.46 |
| CZTS Top cell in tandem structure | 20.98 | 852.9 | 81.95 | 14.67 |
| CZTSe Bottom cell in tandem structure | 20.87 | 474.1 | 72.59 | 7.18 |
| Tandem cell | 20.98 | 1324.82 | 78.2 | 21.7 |

These studies suggest that CZTSSe based tandem structures may provide a way to realize the proposed geometries and thus, the enhanced photovoltaic efficiency. The investigated tandem structure showed 21.7% photovoltaic efficiency for CZTS and CZTSe solar cells under current matching conditions with open circuit voltage ~1.33 V and short circuit current density ~20.98 mA $cm^{-2}$.

## Conclusion

Kesterite CZTS/Se solar cells are investigated using one dimensional SCAPS 1D solar cell simulator. The possible single junctions and tandem structures are simulated for their optimized photovoltaic performance. The proposed tandem absorber cell structures show the potential to realize the enhanced efficiency of kesterite based solar cells beyond the reported maximum efficiency ~12.7 % [4]. The present studies showed the optimal efficiency ~22 % photovoltaic efficiency for tandem CZTS,Se structures. This present analysis demonstrates the possibility of approaching higher efficiency beyond the maximum with present set of tandem structure. Further improvement in the efficiency of tandem cells is possible by introducing wider band gap chalcogenide such as CZGS ($Cu_2ZnGeS_4$) in tandem geometries.

**Acknowledgement**

The authors would like to thank Marc Burgelman and their team for providing SCAPS simulator and financial support from Department of Science & Technology (DST), Government of India, through grants DST/INT/Mexico/P-02/2016.